\documentclass[12pt]{article}
\usepackage{epsfig,amsfonts,amssymb}
\usepackage{cite}
\input epsf.sty
\usepackage{cite}

\def\ZZZ{{\hbox{ Z\kern-1.6mm Z}}}
\def\RRR{{\hbox{ R\kern-2.4mm R}}}
\def\CCC{{\hbox{ C\kern-2.0mm C}}}
\def\zzz{{\hbox{z\kern-1mm z}}}

\newcommand{\qeq}{{\hbox{=\kern-2.3mm ? \kern.5mm }}}
\renewcommand{\qeq}{=}

\newcommand{\wt}{\widetilde}
\newcommand{\wh}{\widehat}

\newcommand{\RR}{{\cal R}}
\newcommand{\NN}{{\cal N}}

\newcommand{\be}{\begin{equation}}
\newcommand{\ee}{\end{equation}}
\newcommand{\ben}{\begin{eqnarray}\displaystyle}
\newcommand{\een}{\end{eqnarray}}

\newcommand{\bea}[1]{\begin{eqnarray}\label{#1} }
\newcommand{\eea}{\end{eqnarray}}

\newcommand{\refb}[1]{(\ref{#1})}

\newcommand{\sectiono}[1]{\section{#1}\setcounter{equation}{0}}

\def\one{{\hbox{ 1\kern-.8mm l}}}
\def\zero{{\hbox{ 0\kern-1.5mm 0}}}

\begin{document}

\begin{center}
{\Large \bf
Duality Orbits, Dyon Spectrum and Gauge Theory Limit of
Heterotic String Theory on $T^6$
}

\end{center}

\vskip .6cm
\medskip

\vspace*{4.0ex}

\centerline{\large \rm  Shamik Banerjee and Ashoke Sen}

\vspace*{4.0ex}

\centerline{\large \rm }

\vspace*{4.0ex}

\centerline{\large \it Harish-Chandra Research Institute}

\centerline{\large \it  Chhatnag Road, Jhusi,
Allahabad 211019, INDIA}

\vspace*{1.0ex}
\centerline{E-mail:  bshamik, sen@mri.ernet.in}

\vspace*{5.0ex}

\centerline{\bf Abstract} \bigskip

For heterotic string theory compactified on $T^6$, we derive the
complete set of T-duality invariants which characterize a pair of
charge vectors $(Q,P)$ labelling the electric and magnetic charges
of the dyon. Using this we can identify the complete set of dyons to
which the previously derived degeneracy formula can be extended.
By going near special points in the moduli space of the theory
we derive the spectrum of quarter BPS dyons in $\NN=4$
supersymmetric gauge theory with simply laced gauge groups.
The results are in agreement with those derived from field theory
analysis.

\vfill \eject

\baselineskip=18pt

\tableofcontents

\sectiono{Introduction} \label{s0}

We now have a good understanding
of the exact spectrum of a class of
quarter BPS dyons in a variety of $\NN=4$
supersymmetric string 
theories\cite{9607026,0412287,0505094,0506249,0508174,
0510147,0602254,
0603066,0605210,0607155,0609109,0612011,0702141,
0702150,0705.1433,0705.3874,0706.2363,0708.1270}. 
Explicit computation of the 
spectrum was carried out for a special class of 
charge vectors in a specific region of the
moduli space. 
Using the various duality invariances of the theory we can extend
the results to various other charge vectors in
various other regions in the moduli space.
However in order to do this we need to find out the duality
orbits of the charge vectors for which the spectrum has been
computed. This is one of the goals of this paper. Throughout
this paper we shall focus on a particular $\NN=4$ supersymmetric
string theory -- heterotic string theory compactified on a six
dimensional torus $T^6$. 

A duality transformation typically acts on the charges as well as
the moduli. Thus using duality invariance we can relate the
degeneracy of a given state at one point of the moduli
space to that of a different state, carrying different set of charges,
at another point of the moduli space. For BPS states however
the degeneracy -- or more precisely an appropriate index 
measuring the number of bosonic supermultiplets
minus the number of fermionic supermultiplets for a given set
of charges
-- is
invariant under changes in the moduli unless we cross a wall
of marginal stability on which the state under consideration becomes
marginally unstable. Thus for BPS states, instead of having to
describe the spectrum as a function of the continuous moduli
parameters we only need to specify it  
in different domains bounded by walls
of marginal stability\cite{0702141,0706.2363,0708.1270}. 
It  turns out that a T-duality transformation
takes a point inside one such domain to another point inside the
same domain in a sense described precisely in 
\cite{0702141,0708.1270}. Thus
once we have calculated the spectrum in one domain for a given
charge, T-duality symmetry can be used to find the spectrum in the
same domain for all other charges related to the initial charge
by a T-duality transformation. For this reason it is important to
understand under what condition two different charges are related to
each other by a T-duality transformation, \i.e.\ to
classify the T-duality orbits.
S-duality transformation, on the other hand, takes a point inside one
domain to a point in another domain. Thus once we have calculated
the spectrum in one domain, S-duality transformation allows us
to calculate the spectrum in other domains.

Our results for the T-duality orbit of charges
can be summarized as follows.
Since heterotic string theory on $T^6$ has a gauge group of rank
28, a typical state is characterized by a 28 dimensional electric charge
vector $Q$ and a 28 dimensional magnetic charge vector $P$, each
taking values on the Narain lattice $\Lambda$
of signature (6,22). 
We shall take $Q$ and $P$ to
be primitive vectors of the lattice;
if not we can express them as integer multiples of primitive
vectors and apply our analysis to these primitive vectors,
treating the integer factors as additional T-duality invariants.
Let $Q_i$
and $P_i$ denote the components of $Q$ and $P$ along some
basis of primitive vectors of the lattice $\Lambda$ 
and $L_{ij}$ denote
the natural metric of signature (6,22) under which the lattice
is even and self-dual. Then the complete set of T-duality invariants
are as follows. First of all we have the invariants of the
continuous T-duality group:
\be \label{econtint}
Q^2 = Q^T L Q, \qquad P^2 = P^T L P, \qquad Q\cdot P 
=Q^T L P\, .
\ee  
Next we have the combination\cite{askitas,0702150}
\be \label{egcdint}
r(Q,P) = \hbox{g.c.d.}\{ Q_i P_j - Q_j P_i, \quad 1\le i,j\le 28\}\, .
\ee 
Finally we have 
\be \label{eu1}
u_1(Q,P) = \alpha \cdot P \, \, \hbox{mod $r(Q,P)$}, \qquad 
\alpha\in \Lambda, \quad \alpha \cdot Q=1\, .
\ee
$u_1(Q,P)$ can be shown to be independent of the choice
of $\alpha\in\Lambda$.
One finds first of all that each of the five combinations
$Q^2$, $P^2$, $Q\cdot P$, $r(Q,P)$ and $u_1(Q,P)$ is
invariant under T-duality transformation. Furhermore two
pairs $(Q,P)$ and $(Q',P')$ having the same set of
invariants can be transformed to each other by a T-duality
transformation. Thus a necessary and sufficient condition for
two pairs of charge vectors $(Q,P)$ and $(Q',P')$ to be
related via a T-duality transformation is that all the five
invariants are identical for the two pairs.

The computation of \cite{0605210} of the 
spectrum of quarter BPS states in heterotic string theory on
$T^6$ has been carried out for a special class of charge vectors
for which $r(Q,P)=1$, and $Q^2$, $P^2$ and $Q\cdot P$ are
arbitrary. The invariant $u_1(Q,P)$ is trivially 0 for states with
$r(Q,P)=1$. 
Let us denote the calculated index by $f(Q^2,P^2,Q\cdot P)$.
Then T-duality invariance tells us that  
for all states with $r(Q,P)=1$ the index is given by the
same function $f(Q^2,P^2,Q\cdot P)$ in the domain
of the moduli space in which the original calculation was
performed. Since S-duality maps states with $r(Q,P)=1$ to states
with $r(Q,P)=1$, but maps the original domain to other domains,
S-duality invariance allows 
us to extend the result
to all states with $r(Q,P)=1$ in all domains of the moduli space.

Since at special points in the moduli space of heterotic string
theory on $T^6$ we can get $\NN=4$ supersymmetric
gauge theories with simply laced gauge 
groups\cite{narain,nsw} in the low energy limit, 
we can use the
dyon spectrum of string theory to extract information about
the dyon spectrum of $\NN=4$ supersymmetric gauge theories.
For this we need to work near the point in the moduli space
where we have enhanced gauge symmetry. Slightly away from this
point we have the non-abelian part of the gauge symmetry
spontaneously broken at a scale small compared to the string
scale, and the spectrum of string theory contains quarter BPS
dyons whose masses are of the order of the symmetry
breaking scale. These dyons can be identified as dyons in the
$\NN=4$ supersymmetric
gauge theory. 
Thus the knowledge of the quarter 
BPS dyon spectrum in heterotic string
theory on $T^6$ gives us information about the quarter BPS
dyon spectrum
in all $\NN=4$ supersymmetric gauge theories which can be
obtained from the heterotic string theory on $T^6$.
This method has been used in \cite{0708.3715} 
to compute the spectrum of
a class of quarter BPS states in $\NN=4$ supersymmetric SU(3)
gauge theory. 

Since the result for the quarter BPS dyon 
spectrum in heterotic string theory on $T^6$
is known only for the states with $r(Q,P)=1$, we can
use this information 
to compute the index of only a subset of dyons in
$\NN=4$ super Yang-Mills theory with simply laced gauge
groups. For this subset of states
the result for the index can be stated in a simple manner,
-- we find that the index 
is non-zero only for those charges which can be
embedded in the root lattice of an SU(3) subalgebra. Thus these
states fall within the class of states analyzed 
in \cite{0708.3715} and
can be represented as arising from a 3-string junction with the
three external strings ending on three parallel 
D3-branes\cite{9712211}. This result for general
$\NN=4$ supersymmetric gauge theories
is in agreement with previous
results obtained either by direct analysis in gauge 
theory\cite{0005275,0609055} or by the
analysis of the spectrum of string network on a system of
D3-branes\cite{9804160}.\footnote{Different
aspects of dyon spectrum in $\NN=4$ supersymmetric gauge
theories have been discussed in \cite{appear}.}

Some related issues have been addressed in \cite{0711.4671}.

\sectiono{T-duality orbits of dyon charges in heterotic string theory
on $T^6$}  \label{s1}

We consider heterotic string theory compactified on $T^6$. In
this case a general dyon is characterized by its electric and
magnetic charge vectors $(Q,P)$ where $Q$ and $P$ are 28
dimensional charge vectors taking values in the Narain lattice
$\Lambda$\cite{narain}. We shall express $Q$ and $P$ as 
linear combinations of a primitive basis of
lattice vectors so that the coefficients 
$Q_i$ and $P_i$ are integers. 
There is a natural metric $L$
of signature (6,22) on $\Lambda$ under which the lattice is
even and self-dual. The discrete T-duality transformations
of the theory take the form
\be \label{edist}
Q\to \Omega Q, \qquad P\to \Omega P\, ,
\ee
where $\Omega$ is a $28\times 28$ matrix that preserves the
metric $L$ and the Narain lattice $\Lambda$
\be \label{eomega}
\Omega^T L \Omega = L, \qquad \Omega \Lambda = \Lambda\, .
\ee
Since $\Omega$ must map an arbitrary
integer valued vector to another integer valued vector, the
elements of $\Omega$ must be integers.

We shall assume from the
beginning that $Q$ and $P$ are primitive elements of the
lattice.\footnote{If this is not the case then 
the gcd $a_1$ of all the elements
of $Q$ and the gcd $a_2$
of all the elements of $P$ will be
separately invariant under discrete T-duality transformation.
We can factor these out as $Q=a_1 \, \overline{Q}$,
$P=a_2 \, \overline{P}$ with $a_1,a_2\in\ZZZ$,
$\overline{Q},\overline{P}\in\Lambda$,
and then apply our analysis on the
resulting primitive elements $\overline{Q}$ and 
$\overline{P}$.}
Our goal is to find the T-duality invariants which characterize
the pair of charge vectors $(Q,P)$. 
First of all we have the continuous T-duality
invariants
\be \label{econt}
Q^2 = Q^T L Q, \qquad P^2 = P^T L P, \qquad Q\cdot P 
=Q^T L P\, .
\ee  
Besides these we can introduce some additional invariants
as follows. Consider the combination\cite{askitas,0702150}
\be \label{egcd}
r(Q,P) = \hbox{g.c.d.}\{ Q_i P_j - Q_j P_i, \quad 1\le i,j\le 28\}\, .
\ee 
We shall first show that $r(Q,P)$ is independent of the choice
of basis in which we expand $Q$ and $P$. For this we note that
the component form of $Q$ and $P$ in
a different choice of basis will be related to the ones given above by
multiplication by
a matrix $S$ with integer elements and unit determinant so
that the elements of $S^{-1}$ are also
integers. Thus in this new basis $r$ will be given by
\be \label{em1}
r(SQ,SP) = \hbox{gcd} \left\{S_{ik} S_{jl} (Q_k P_l - Q_l P_k),
\quad 1\le i,j\le 28\right\}\, .
\ee
Since $S_{ik}$ are integers, eq.\refb{em1} shows that
$r(SQ,SP)$ must be divisible by
$r(Q,P)$. 
Applying the
$S^{-1}$ transformation on $(S Q, S P)$,
and noting that $S^{-1}$ also has integer elements,
we can show that $r(Q,P)$ must be divisible
by $r(SQ,SP)$.
Thus we have
\be \label{em2}
r(Q, P) = r(S Q, S P)\, ,
\ee
\i.e. $r(Q,P)$ is independent of the choice of basis used to describe
the vectors $(Q,P)$. As a special case where we restrict $S$ to
T-duality transformation matrices $\Omega$, we find
\be \label{e2}
r(Q, P) = r(\Omega Q, \Omega P)\, .
\ee
Thus
$r(Q,P)$ is invariant under a T-duality transformation.

Another set of T-duality invariants may be constructed as
follows. Let $\alpha,\beta\in \Lambda$ satisfy 
\be \label{e3}
\alpha\cdot Q=1, \qquad \beta \cdot P=1\, .
\ee 
Since $Q$ and $P$ are primitive and the 
lattice is self-dual one can
always find such $\alpha$, $\beta$. Then we define
\be \label{e4}
u_1(Q,P) = \alpha \cdot P \, \, \hbox{mod $r(Q,P)$}, 
\qquad u_2(Q,P)=\beta \cdot Q\, \, \hbox{mod $r(Q,P)$}\, .
\ee
One can show that\cite{askitas}
\begin{enumerate}
\item $u_1$ and $u_2$ are independent of the choice of 
$\alpha$, $\beta$.
\item $u_1$ and $u_2$ are T-duality invariants.
\item $u_2$ is determined uniquely in terms of $u_1$.
\end{enumerate}
The proof of these statements goes as follows.
To prove that $u_1$ is independent of the choice of
$\alpha$ we note that since $Q$ is a primitive vector we
can choose a basis of lattice vectors so that the first element
of the basis is $Q$ itself. Then in this basis\footnote{Note
that in this basis the metric takes a complicated form, {\it e.g.}
the 11 component of the metric must be equal to $Q^2$.
However all components of the metric are still integers since
the inner product between two arbitrary integer valued vectors
-- representing a pair of elements of the lattice -- must be integer.}
\be \label{eqp}
Q=\pmatrix{1\cr 0\cr \cdot\cr \cdot \cr 0},
\qquad P = \pmatrix{P_1\cr P_2\cr \cdot\cr\cdot\cr P_{28}}\, ,
\ee
and we
have
\be \label{e5}
r(Q,P) = \hbox{gcd}(P_2,\cdots P_{28})\, .
\ee
Now suppose $\alpha_1$ and $\alpha_2$ are two vectors which
satisfy $Q\cdot\alpha_1=Q\cdot\alpha_2=1$. Then $(\alpha_1
-\alpha_2)\cdot Q=0$, and hence we have
\be \label{e6}
(\alpha_1-\alpha_2)\cdot P = (\alpha_1-\alpha_2)\cdot
(P - P_1 Q) = (\alpha_1-\alpha_2)\cdot
\pmatrix{0\cr P_2 \cr P_3\cr \cdot\cr \cdot\cr P_{28}}\, .
\ee
Eq.\refb{e5} shows that the right hand side of \refb{e6} is
divisible by $r$. Thus $\alpha_1\cdot P=\alpha_2\cdot P$
modulo $r$. This shows that $u_1$ defined through
\refb{e4} is independent of the choice of $\alpha$. A similar
analysis shows that $u_2$ defined in \refb{e4}
is independent of the choice of $\beta$. {}From now on
all equalities involving $u_1(Q,P)$ and $u_2(Q,P)$ will be
understood to hold modulo $r(Q,P)$ although we shall not
always mention it explicitly. 

$T$-duality invariance of $u_1$ follows from the fact that
if $\alpha\cdot Q=1$ then $\Omega\alpha \cdot \Omega Q=1$.
Thus
\be \label{e7}
u_1(\Omega Q, \Omega P) = \Omega\alpha \cdot \Omega P
 \, \, \hbox{mod $r(\Omega Q,\Omega P)$}
= \alpha \cdot P \, \, \hbox{mod $r(Q,P)$} = u_1(Q,P)
\, .
\ee
A similar analysis shows the T-duality invariance of $u_2$.

To show that $u_2$ is determined in terms of $u_1$ and vice
versa we first note that for the choice of $(Q,P)$ given in
\refb{eqp}, we have
\be \label{eau1}
u_1(Q,P) = \alpha \cdot P = \alpha \cdot (P- P_1 Q)
+ P_1\, \alpha\cdot Q = P_1 \, \, \hbox{mod $r(Q,P)$}
\ee
since $P-P_1 Q$ is divisible by $r$ due to eqs.\refb{eqp},
\refb{e5}, and $\alpha\cdot Q=1$. On the other hand we
have 
\be \label{eau2}
1 = \beta\cdot P = \{\beta \cdot (P-P_1 Q) + P_1 \beta\cdot Q
\} 
= u_1(Q,P) u_2(Q,P) \,\, \hbox{mod $r(Q,P)$}\, ,
\ee
since $(P-P_1 Q)=0$ modulo
$r$, $P_1=u_1$, and $\beta\cdot Q = u_2$.
Thus we have
\be \label{e8}
u_1(Q,P) \, u_2(Q,P) = 1 \, \, \hbox{mod $r(Q,P)$}\, .
\ee
This shows that neither $u_1$ nor $u_2$ shares a common
factor with $r$. We shall now show that \refb{e8} also
determines $u_2$ uniquely in terms of $u_1$. To prove
this assume the contrary, that there exists another number
$v_2$ satisfying $u_1\, v_2 = 1\, \, 
\hbox{mod $r(Q,P)$}$. Then
we have
\be \label{e9}
 u_1(Q,P)\, \left(u_2(Q,P)  - v_2(Q,P)\right) = 0
\, \, 
\hbox{mod $r(Q,P)$}\, .
\ee
Since $u_1$ has no common factor with $r$, this shows that
$v_2 = u_2$ modulo $r$. Hence $u_2$ is determined
in terms of $u_1$ modulo $r$.

Thus we have so far identified five separate T-duality  invariants
characterizing the pair of vectors $(Q,P)$: $Q^2$, 
$P^2$, $Q\cdot P$, $r(Q,P)$ and $u_1(Q,P)$. We shall now
show that these are sufficient to characterize a T-duality orbit, \i.e.\
given any two pairs $(Q,P)$ and $(Q',P')$ with the same set of
invariants they are related by a T-duality transformation.
We begin by defining\footnote{This procedure breaks down for
$Q^2=0$, but as long as $P^2\ne 0$ we can carry out our analysis
by reversing the roles of $Q$ and $P$. If both $Q^2$ and $P^2$
vanish then our analysis does not apply.  However a different proof
given in 
\S\ref{snew} applies to this case as well.}
\be \label{e10}
\wh P = Q^2 P - Q\cdot P \, Q\, ,
\ee
and 
\be \label{e11}
\wt P = {1\over K} \, \wh P\, , \qquad K \equiv \hbox{gcd}\{\wh P_1,
\cdots \wh P_{28}\}\, .
\ee
By construction $\wt P$ is a primitive vector of the lattice
satisfying
\be \label{e12}
Q\cdot \wt P = 0\, .
\ee
We shall now use the result of 
\cite{askitas} that the T-duality orbit of a pair of primitive
vectors $(Q,\wt P)$
satisfying $Q\cdot \wt P=0$ is characterized completely by
the invariants $Q^2$, $\wt P^2$, $r(Q,\wt P)$ and
$u_1(Q,\wt P)$. A proof of this statement has been
reviewed in appendix \ref{sa}. Given this, 
we shall show that the five invariants $Q^2$, $P^2$,
$Q\cdot P$, $r(Q,P)$ and $u_1(Q,P)$ completely
characterize the duality
orbits of an arbitrary pair of charge vectors $(Q,P)$. The steps
involved in the proof are as follows:
\begin{enumerate}
\item We shall first show that the quantities 
$\wt P^2$, $r(Q,\wt P)$,
$u_1(Q,\wt P)$ and the 
constant $K$ appearing in \refb{e11}
are determined completely in terms of
$Q^2$, $P^2$,
$Q\cdot P$, $r(Q,P)$ and $u_1(Q,P)$ via the relaions
\ben\label{erelinv}
&& K = r(Q,P) \, \hbox{gcd} \, \left\{
 (u_1(Q,P) Q^2 - Q\cdot P)/ r(Q,P), Q^2\right\}, \nonumber \\
&& r(Q, \wt P) = Q^2 r(Q,P) / K, \quad
u_1(Q,\wt P) = {1\over K}\, (u_1(Q,P) Q^2 - Q\cdot P)
\quad \hbox{mod $r(Q,\wt P)$}\, ,
\nonumber \\
&&
\wt P^2 = {1\over K^2} Q^2 (Q^2 P^2 - (Q\cdot P)^2)
 \, . 
\een
The last equation follows trivially from the definition of
$\wt P$.
To prove the other relations we again use the form of $(Q,P)$
given in \refb{eqp}. We have
\be \label{ephat}
\wh P = Q^2 \, P - Q\cdot P \, Q
= Q^2 (P - P_1 Q) - Q\cdot (P - P_1 Q) \, Q
= r(Q,P) \{ Q^2 \gamma - Q\cdot \gamma \, Q)\, ,
\ee
where
\be \label{edefgamma}
 \gamma = {1\over r(Q,P)} (P - P_1 Q) = {1\over r(Q,P)}
\pmatrix{0\cr P_2\cr \cdot\cr\cdot\cr P_{28}}\, .
\ee
$\gamma$ has integer elements due to \refb{e5}.
The same equation tells us that
\be \label{eau4}
\hbox{gcd}(\gamma_2, \cdots \gamma_{28}) = 1\, .
\ee
Expressing \refb{ephat} as
\be \label{eau5}
\wh P = r(Q,P)\pmatrix{-Q\cdot\gamma\cr Q^2 \gamma_2\cr
\cdot\cr \cdot \cr Q^2 \gamma_{28}}\, ,
\ee
and using \refb{eau4}
we see that $K$ defined in \refb{e11} is given by
\be \label{eau6}
K = r(Q,P)\, \hbox{gcd}(-Q\cdot \gamma, Q^2)\, .
\ee
Using \refb{edefgamma} and that $P_1=u_1(Q,P)$ modulo
$r(Q,P)$ we may express \refb{eau6} as
\be \label{eau7}
K = r(Q,P) \, \hbox{gcd} \, \left\{
 (u_1(Q,P) Q^2 - Q\cdot P)/ r(Q,P), Q^2\right\}\, .
 \ee
 This establishes the first equation in \refb{erelinv}.
 Note that a shift in $u_1$ by $r(Q,P)$ does not change the
 value of $K$. Thus $K$ given in \refb{eau7} is independent 
 of which particular representative we use for $u_1(Q,P)$.
 
 To derive an expression for $r(Q,\wt P)$ we note from
 the form of $Q$ given in \refb{eqp}, the form of
 $\wh P$ given in \refb{eau5}, and \refb{eau4} that
 \be \label{eau8}
 r(Q,\wh P)=
 \hbox{gcd}\{Q_i \wh P_j - Q_j \wh P_i, \,\, 1\le i,j\le 28\}
 = r(Q,P) \, Q^2\, .
 \ee
Since $\wt P= \wh P / K$ we have
\be \label{eau9}
r(Q,\wt P) = Q^2 \, r(Q,P) / K\, .
\ee
This establishes the second equation in \refb{erelinv}.
Finally to calculate $u_1(Q, \wt P)$ we pick the vector $\alpha$
for which $\alpha\cdot Q=1$, and express $u_1(Q,\wt P)$ as
\be \label{eau10}
u_1(Q,\wt P) = \alpha \cdot \wt P
= {1\over K} \left(Q^2 \alpha\cdot P - Q\cdot P \, \alpha\cdot Q
\right) = {1\over K} \left( Q^2 u_1(Q,P) - Q\cdot P\right)\, .
\ee
This establishes the third equation in \refb{erelinv}. Note that
under a shift of $u_1(Q,P)$ by $r(Q,P)$, the expression 
for $u_1(Q,\wt P)$ given above shifts by $r(Q,\wt P)$. Thus
$u_1(Q,\wt P)$ given above is determined unambiguously
modulo $r(Q,\wt P)$.

\item Now suppose we have  two pairs $(Q,P)$ and $(Q',P')$ with
the same set of invariants:
\be \label{e13}
Q^2=Q^{\prime 2}, \quad P^2 = P^{\prime 2}, \quad
Q\cdot P = Q'\cdot P', \quad r(Q,P) = r(Q',P'), \quad
u_1(Q,P)=u_1(Q',P')\, .
\ee
Let us define $\wh P'$, $K'$ and $\wt P'$
as in \refb{e10}, \refb{e11} with $(Q,P)$ replaced by $(Q',P')$
so that $Q'\cdot\wt P'=0$.
Then by eq.\refb{erelinv}, 
its analog with $(Q,P)$ replaced by $(Q',P')$,
and eq.\refb{e13}, we have
\be \label{e14}
Q^2=Q^{\prime 2}, \quad K'=K, \quad \wt P^2 = \wt P^{\prime 2}, 
\quad
 r(Q,\wt P) = 
r(Q',\wt P'), \quad
u_1(Q,\wt P)=u_1(Q',\wt P')\, .
\ee
Thus by the result of \cite{askitas}, reviewed in appendix \ref{sa},
$(Q,\wt P)$ and $(Q',\wt P')$
must be related to each other by a T-duality transformation
$\Omega$:
\be \label{e15}
Q'=\Omega Q, \qquad \wt P' = \Omega \wt P\, .
\ee
It follows from this that
\be \label{e16}
\wh P' = \Omega\wh P, \quad \longrightarrow
\quad P'=\Omega P\, .
\ee
Thus $(Q,P)$ and $(Q',P')$ are related by the duality
transformation $\Omega$. 

\end{enumerate}

This establishes that the T-duality orbits of pairs of charge
vectors $(Q,P)$ are completely characterized by the invariants
$Q^2$, $P^2$, $Q\cdot P$, $r(Q,P)$ and $u_1(Q,P)$. Two
pairs of charge vectors, having the same values of all the
invariants, can be related to each other by a T-duality
transformation.

\sectiono{An Alternative Proof} \label{snew}

In this section we shall give a different proof of the results of the
previous section.

We shall begin by giving a physical
interpretation of the discrete T-duality invariants $r(Q,P)$
and $u_1(Q,P)$. Let $E$ denote the two dimensional 
vector space spanned by the vectors
$Q$ and $P$, and  $\Lambda'=E\cap\Lambda$ denote the two
dimensional lattice containing
the points of the Narain lattice in $E$. 
Let $(e_1, e_2)$ denote a pair of primitive basis elements of the
lattice $\Lambda'$. 
Since $Q$ is a primitive vector, we can always
choose $e_1=Q$.
Then we claim that in this basis 
\be \label{einthis}
Q=e_1, \qquad P = u_1(Q,P)\, e_1 + r(Q,P)\, e_2\, .
\ee
The proof goes as follows. First of all since $(e_1,e_2)$ form a
primitive basis of $\Lambda'$, by a standard result\cite{martinet}
one can show that
$(e_1,e_2)$ can be chosen as the first two elements of a
primitive basis of the full lattice $\Lambda$. In such a basis
$Q_1=1$, $P_1=u_1$, $P_2=r$ and all the other componets
of $Q$ and $P$ vanish. Thus we have 
gcd~$\{Q_iP_j-Q_jP_i\}=r$ as required. Furthermore,
it is clear from \refb{einthis} that if $\alpha\cdot Q=1$ then 
$\alpha\cdot P=u_1$ modulo $r$
as required by the definition of
$u_1$. Finally, we see that a different choice of $e_2$ that
preserves the primitivity of the basis $(e_1,e_2)$ is
related to the original choice by $e_2\to e_2+ s\, e_1$ for
some integer $s$. Under such a transformation $u_1$
defined through \refb{einthis} is shifted by a multiple of
$r$. Thus $u_1$ defined through \refb{einthis} is
unambiguous modulo $r$ as required. We shall choose $e_2$
such that $u_1$ appearing in \refb{einthis} lies between
0 and $r-1$.

Eq.\refb{einthis} provides a physical interpretation of
$u_1$ and $r$ in terms of the components of $Q$ and $P$ along
a primitive basis of the Narain lattice in the plane
spanned by $Q$ and $P$. As a
consequence of \refb{einthis} we have
\ben \label{ee1e2}
&& e_1^2 = Q^2, \qquad e_2^2 =  \left\{P^2
+ u_1(Q,P)^2 Q^2 - 2 u_1(Q,P) Q\cdot P\right\}/
r(Q,P)^2, \nonumber \\
&&
e_1\cdot e_2 =  \left\{Q\cdot P - u_1(Q,P)
Q^2\right\} /  r(Q,P) \, .
\een
Now take a different pair of charges $(Q',P')$ with the same
invariants, {\it e.g.} satisfying \refb{e13}, and define $(e_1',e_2')$
as in \refb{einthis} with  $(Q,P)$ replaced by $(Q',P')$. Then
as a consequence of \refb{e13} and \refb{ee1e2} we have
\be \label{e3.2}
e_1^2 = (e_1')^2, \qquad e_2^2 = (e_2')^2, \qquad 
e_1\cdot e_2 = e_1'\cdot e_2'\, .
\ee
Thus the lattices generated by $(e_1,e_2)$ and $(e_1',e_2')$
can be regarded as different primitive embeddings into $\Lambda$
of  an abstract even
lattice of rank two  with a given metric. 
We now use the result of \cite{james,loji,nikulin} that
an even lattice  of
signature $(m,n)$ has a unique primitive embedding
in an even self-dual 
lattice $\Lambda$ of signature $(p,q)$ up to a T-duality
transformation if
$m+n\le \hbox{min}(p,q)-1$. Setting $m+n=2$ and $(p,q)=(6,22)$
we see that the required condition is satisfied and hence 
$(e_1,e_2)$ must be related to $(e_1',e_2')$ by a T-duality
transformation:
\be \label{e3.3}
e_1'=\Omega e_1, \qquad e_2'=\Omega e_2\, .
\ee
Eq.\refb{einthis} and its analog with $(Q,P)\longrightarrow (Q',P')$,
$(e_1,e_2)\longrightarrow(e_1',e_2')$ then tells us that
\be \label{e3.4}
Q'=\Omega Q, \qquad P'=\Omega P\, .
\ee
This is the desired result.

One interesting question is: for a given set of values of $Q^2$, $P^2$,
$Q\cdot P$ and $r$, what is the maximum number of possible
orbits? This is given by the maximum number of allowed values
of $u_1$. Since $u_1$ and $r$ cannot share a common factor, the
number is bounded from above by the number of positive 
integers below
$(r-1)$ with no common factor with $r$. This in turn is
given by
\be \label{eprimefac}
r\, \times \, \prod_{\hbox{primes $p$,  $p|r$}} 
\left( 1 - {1\over p}\right)\, .
\ee

\sectiono{Predictions for gauge theory} \label{s3}

At special points in the moduli space heterotic string theory
on $T^6$ has enhanced gauge symmetry. As we move away from this
point the gauge symmetry gets spontaneously broken, with the moduli
fields
describing deformations away from the enhanced symmetry point
playing the role of the Higgs field.
When the deformation parameter is small the scale of gauge symmetry
breaking is small compared to the string scale and the
theory contains massive
states with mass of the order of the gauge symmetry breaking scale
and small compared to the string scale. These states
can be identified as the states of the  spontaneously broken gauge theory.
Thus if we know the spectrum of the string theory, we can determine
the spectrum of spontaneously broken gauge theory. In particular the
known spectrum of quarter BPS dyons in string theory should give
us information about the spectrum of quarter BPS dyons in $\NN=4$
supersymmetric Yang-Mills theory.

The dyon charges in a gauge theory of 
rank $n$ are labelled by a pair
of $n$-dimensional vectors $(q,p)$  in the root lattice of the
gauge algebra. If we choose a set of $n$ simple roots as the
basis of the root lattice  then the components $q_a$ and $p_a$
will label the coefficients of the simple roots in an
expansion of the charge vectors in this basis. 
When the root lattice is embedded in the Narain lattice the
vectors $(q,p)$ correpond to a pair of vectors $(Q,P)$ in the 
Narain lattice, and the metric $L$ on the Narain lattice,
restricted to the root lattice, gives the negative of the
Cartan metric. 
Denoting
by $\circ$ the inner product with respect to the Cartan metric, we have
\be \label{es1}
q^2\equiv q\circ q=-Q^2, \qquad 
p^2 \equiv p\circ p = - P^2, \qquad q\circ p = - Q\cdot P\, .
\ee
Since the Cartan metric is positive definite,
we must have $q^2,p^2\ge 0$, $|q\circ p|\le (q^2+p^2)/2$. 
Furthermore quarter BPS dyons
require $q$ and $p$ to be both non-zero and non-parallel. Hence
none of the above inequalities can be saturated. 
This translates to the
following conditions on $Q$, $P$:
\be \label{es4}
Q^2 < 0, \qquad P^2 < 0\, \qquad |Q\cdot P| < \left(|Q^2|
+|P^2|\right) / 2\, .
\ee
Finally, since the string theory 
dyon spectrum is known only for 
charges $(Q,P)$ with $r(Q,P)=1$ we need to know what this
condition
translates to on the vectors $(q,p)$.
This is done most easily if the Narain lattice admits a
primitive embedding of the root lattice, \i.e.\
if we
can choose the $n$ simple roots of the
root lattice as the first $n$ basis
elements of the full $28$ dimensional
Narain lattice.  
In that case we can easily identify 
$(q,p)$ in the
root lattice
as  a pair of charge vectors $(Q,P)$ in the
Narain lattice where the first $n$ components 
of $Q$ ($P$) are equal
to the components of $q$ ($p$) and the rest of the components of
$Q$, $P$ vanish.  Thus we have
\be \label{es2}
r(Q, P) = \hbox{gcd}\{ q_i p_j - q_j p_i\} \equiv r_{gauge}(q,p)\, .
\ee
The condition $r(Q,P)=1$ then translates to
$r_{gauge}(q,p)=1$. 

Let us now investigate under what condition the Narain
lattice does not admit a primitive embedding of the root lattice.
Let $F$ be the
$n$-dimensional 
vector space spanned by the root lattice, and let $\Lambda'=
F\cap\Lambda$.
Then by a standard result\cite{martinet} one finds that
the root lattice has a primitive embedding 
in the Narain lattice if
$\Lambda'$ does not contain any
element other than the
ones in the root lattice. 
So we need to examine under what condition $\Lambda'$
can contain elements other than the ones in the root lattice.
Now clearly the elements
of $\Lambda'$ must belong to the weight lattice of the
algebra. Furthermore, since Narain lattice is even, any 
element of $\Lambda'$ will be even. Thus we can classify
all possible extra elements of $\Lambda'$ by examining the
possible even elements of the weight lattice outside the root
lattice. For many algebras we have no such element, and hence
in those cases the embedding of the root lattice in the Narain
lattice is necessarily primitive.
Exceptions among the rank $\le 22$ algebras
are $so(16)$, $so(32)$, $su(8)$,
$su(9)$, $su(16)$ and $su(18)$; 
for each of these the weight lattice has even elements
other than those in the root 
lattice\cite{conway}.\footnote{Both for $so(16)$
and $su(9)$, inclusion of the extra even elements of the
weight lattice makes the lattice $F\cap \Lambda$ into the
root lattice of $e_8$. Thus for such embeddings we are actually
counting the dyon spectrum of an $E_8$ gauge theory rather
than $SO(16)$ or $SU(9)$ gauge theory. On the other hand
for $su(8)$ the extra even elements of the weight lattice
makes $F\cap \Lambda$ into the root lattice of $e_7$. Thus
in this case we get an $E_7$ gauge theory.}
Hence in these cases $\Lambda'$ could contain elements
other than the ones in the root lattice, preventing
the root lattice from having a primitive embedding in the
Narain lattice.  But since $\Lambda'$ would have a primitive
embedding in the Narain lattice, 
 if we
choose a basis for $\Lambda'$, and define $q_i$, $p_i$ as
the componets of $q$ and $p$ expanded in this basis, then
\refb{es2} continues to reproduce the value of 
$r(Q,P)$.

With this understanding we can now study the implications of the
known dyon spectrum in $\NN=4$ supersymmetric string theory.
As is well known, for dyons with $r(Q,P)=1$
the dyon spectrum in different parts of the moduli
space can be different. 
The situation is best described in the axion-dilaton moduli space at
fixed values of the other moduli\cite{0702141}.
In particular in the upper 
half plane labelled by the
axion-dilaton field\footnote{From the point of view of the gauge
theory the axion-dilaton moduli correspond to the theta parameter
and the inverse square of the coupling constant.}
$\tau=a+iS$ 
the spectrum jumps across walls of marginal
stability, which are circles or straight lines passing through rational
points on the real 
axis\cite{0702141,0705.3874,0706.2363}. 
These curves do not intersect in the interior
of the upper half plane and divide up the upper half
plane into different domains, each with three vertices lying either at
rational points on the real axis or at $\infty$.
Inside a given domain the index $d(Q,P)$ that counts the number
of bosonic supermultiplets minus the number of fermionic
supermultiplets
remains constant, but
as we move from one domain to another the index changes.
We shall first consider the domain bounded
by a straight line passing through 0, a straight line passing through 1
and a circle passing through 0 and 1, -- the domain called $\RR$ in
\cite{0702141,0708.1270}. This has vertices at 0, 1 and $\infty$.
In this domain the only non-zero values of $d(Q,P)$ for
$Q^2<0$, $P^2<0$ are obtained at $Q^2=P^2=-2$.
For $Q^2=P^2=-2$ the result for $d(Q,P)$ 
is\cite{0605210,0702141,0708.3715}
\be \label{epar1}
d(Q, P) =\cases{0 \quad \hbox{for 
$Q\cdot P\ge 0$} \cr 
j(-1)^{j-1}
\quad \hbox{for $Q\cdot P = -j$, $j>0$}}\, .
\ee
The condition \refb{es4} on $Q\cdot P$ now shows that for
$(Q,P)$ describing the elements of the root lattice, 
non-vanishing index exists only for $Q^2=P^2=-2$,
$Q\cdot P=-1$.
Translated to a condition on the charge vectors in the gauge
theory this gives\footnote{It has been shown in
appendix \ref{sb} that for states with $Q^2=P^2=-2$,
$Q\cdot P=\pm 1$ the condition $r(Q,P)=1$ is
satisfied automatically. Thus we do not need to state this as
a separate condition.}
\be \label{es6}
d_{gauge}(q,p) =\cases{1 \quad \hbox{for $q^2=p^2=2$,
$q\circ p= 1$, $r_{gauge}(q,p)=1$} \cr 
0
\quad \hbox{for other $(q,p)$ with $r_{gauge}(q,p)=1$}}\, .
\ee
This condition in turn implies that $q$ and $-p$ can be regarded as
the simple roots of an
$su(3)$ subalgebra of the full gauge algebra, with the 
Cartan metric of su(3) being equal to the restriction of the
Cartan metric of the full algebra. 
Thus we learn that in the domain $\RR$ the only dyons with
$r_{gauge}(q,p)=1$ and non-vanishing index are the
ones which can be regarded as $SU(3)$ dyons for some 
level one $su(3)$
subalgebra of the gauge algebra, with  $q$ 
and $-p$ 
identified with the simple roots $\alpha$ and $\beta$
of the $su(3)$ algebra.

The index in other domains can be found using the S-duality
 invariance of the theory. An S-duality transformation of the form
 $\tau\to (a\tau+b)/(c\tau +d)$ maps the domain $\RR$ to another
 domain with vertices
 \be \label{es7}
 {a\over c}, \quad {b\over d}, \quad {a+b\over c+d}\, .
 \ee
 Under the same S-duality transformation  the charge vector
 $(q,p)=(\alpha, -\beta)$ gets mapped to
 \be \label{es8}
 (q,p) = (a\alpha - b\beta, c\alpha - d\beta)\, .
 \ee
 It can be easily seen that $r(Q,P)$ remains invariant under
 an $SL(2,\ZZZ)$ transformation:
 \be \label{erqpinv}
 r(Q,P)=r(aQ+bP, cQ+dP), \quad a,b,c,d\in \ZZZ, \quad ad-bc=1\,.
 \ee
 Thus we conclude that in the domain \refb{es7},
 the index of gauge theory dyons  with $r_{gauge}(q,p)=1$
 is given by
 \be \label{es9}
 d_{gauge}(q,p) =\cases{1 \quad \hbox{for $(q,p)=
 (a\alpha - b\beta, c\alpha - d\beta)$,
} \cr 
0
\quad \hbox{otherwise}}\, ,
\ee
with $(\alpha,\beta)$ labelling the simple roots of some
level one su(3) subalgebra of the full gauge algebra.

This general result agrees with the known results for quarter
BPS dyons in   gauge 
theories\cite{9712211,9804160,
9907090,0005275,0609055}.\footnote{In 
codimension $\ge 1$ subspaces
of the moduli space the dyon spectrum, computed in 
some approximation,  has a rich 
structure\cite{9804160,
9907090,0005275,0609055}.
However the index associated with these dyons vanish and these
results are not in contradiction with the spectrum of
string theory.}
In particular in the
representation of $SU(N)$ dyons as string network with ends
on a set of parallel D3-branes, this is a reflection of the fact that
networks with three external strings ending on three D3-barnes
are the only quarter BPS configurations at a generic point 
in the moduli space\cite{9804160}.

\bigskip

{\bf Acknowledgement:} We wish to thank Suvrat Raju for useful 
discussions and for drawing our
attention to the results of \cite{wall}. We would also like to thank
Nabamita Banerjee, 
Justin David and Dileep Jatkar for useful discussions and 
Justin David for comments
on the manuscript.

\appendix

\sectiono{T-duality orbits of pair $(Q,\wt P)$ with $Q\cdot \wt P=0$}
\label{sa}

In this appendix we shall review the proof, given in 
\cite{askitas}, of the fact that for a pair of primitive
lattice vectors $(Q,\wt P)$ with $Q\cdot \wt P=0$, the
invariants $Q^2$, $\wt P^2$, $r(Q,\wt P)$ and $u_1(Q,\wt P)$
completely characterize the T-duality orbit.
We shall choose a basis in which the metric $L$ given by
the
direct sum of six $\sigma_1$'s and two $-L_{E_8}$'s
where $\sigma_1=\pmatrix{0 & 1\cr 1 & 0}$ and
$L_{E_8}$ is the
Cartan metric
of $E_8$.
Using the known result\cite{wall} that in this lattice
any pair of primitive vectors of the same norm can
be related by a T-duality transformation, we can choose the vector
$Q$ to be
\be \label{ea1}
Q = \pmatrix{-1\cr n\cr 0\cr \cdot\cr \cdot\cr 0}\, , \qquad n=-Q^2/2\, .
\ee
After this T-duality transformation the new
$\wt P$ satisfying $Q\cdot \wt P=0$
has the form $\pmatrix{k\cr kn \cr \vec p}$
for some 26 dimensional vector $\vec p$ and some integer $k$.
In general $\vec p$ is not a primitive vector; 
however if $l$ is the gcd of the
elements of $\vec p$ then $\vec p/l$ is a primitive vector.
We can now use the result of \cite{wall} on the vector $\vec p/l$,
belonging to the Narain lattice of signature (5,21), to bring $\wt P$
into the form\footnote{The result of \cite{wall} is valid on a
Lorenzian lattice of signature $(k,k+16)$ if $k\ge 2$.}
\be \label{ea2}
\wt P' = \pmatrix{k\cr kn \cr l\cr lm\cr 0\cr \cdot\cr \cdot\cr 0}\, ,
\qquad  m={\vec p^2\over 2 \, l^2}
\in\ZZZ\, , \ee
via a T-duality transformation acting on the last 26 elements
that does not affect the form
of $Q$. Furthermore since $\wt P$ 
is a primitive vector so is
$\wt P'$, and we
have
\be \label{ea2.5}
\hbox{gcd}(k,l)=1\, .
\ee
It follows from \refb{ea2.5} and
the definitions of $r(Q,P)$ and $u_1(Q,P)$
given in \refb{egcd}, \refb{e4} that 
\be \label{ea3}
r(Q,\wt P) = r(Q,\wt P')=
\hbox{gcd}(l, 2kn) = \hbox{gcd}(l,2n),
\qquad u_1(Q,\wt P) = u_1(Q,\wt P')=-k\,.
\ee
In arriving at \refb{ea3} we have chosen $\alpha
=\pmatrix{0\cr -1\cr 0\cr \cdot\cr \cdot \cr 0}$.

We now consider the T-duality transformation generated by
\be \label{ea4}
\Omega' = \pmatrix{1 & 0 & 0 & 0 & -1 & 0 &\cr
0 & 1 & 0 & 0 & -n & 0 & \cr 0 & 0 & 1 & 0 & 0 & 0 &\cr
0 & 0 & 0 & 1 & 0 & 0 & \cr n & 1 & 0 & 0 & -n & 1 &\cr
0 & 0 & 0 & 0 & 1 & 0& \cr &&&&&& I_{22}}\, .
\ee
This leaves the charge vector $Q$ invariant, but transforms
$\wt P'$ to
\be \label{ea5}
\wt P'' = \pmatrix{k \cr kn \cr l\cr lm \cr 2 k n\cr 0\cr \cdot
\cr \cdot \cr 0}\equiv \pmatrix{k \cr kn \cr \vec p''}\, .
\ee
We now regard the vector $\vec p''$ as an element of Narain lattice
of signature $(5,21)$. The gcd of all the elements of $\vec p''$
is given by
\be \label{ea6}
\hbox{gcd}(l, 2kn) = \hbox{gcd}(l,2n) = r(Q,\wt P)\, ,
\ee
using \refb{ea2.5}
and \refb{ea3}. Thus  $\vec p''$ is $r(Q,\wt P)$ times a primitive
lattice vector. Hence we can again use the result of \cite{wall}
to show that by a T-duality transformation 
acting on the last 26 elements of the charge
vector, $\vec p''$ can be brought into the form
\be \label{ea7}
\pmatrix{r(Q,\wt P)\cr r(Q,\wt P) \, a \cr 0\cr \cdot\cr \cdot\cr 0}
\, , \qquad a = {\vec p^{\prime \prime 2} \over 2 r(Q,P)^2}\, .
\ee
This does not change the form of $Q$.
Thus at this stage
we have brought $(Q,\wt P)$ to the form
\be \label{ea8}
Q = \pmatrix{-1\cr n\cr 0\cr \cdot\cr \cdot\cr 0}\, , 
\quad \wt P''' = \pmatrix{k \cr kn \cr  r(Q,\wt P)\cr r(Q,\wt P) 
\, a \cr 0\cr \cdot\cr \cdot\cr 0}\, .
\ee

Finally we apply another T-duality transformation generated
by the matrix
\be \label{ea9}
\Omega'' = \pmatrix{1 & 0 & -q & 0&\cr 0 & 1 & - nq & 0&\cr
nq & q & -nq^2 & 1&\cr 0 & 0 & 1 & 0 & \cr &&&& I_{24}}\,,
\ee
with $q$ is an integer to be specified below. This leaves
$Q$ unchanged but brings $\wt P'''$ to the form
\be \label{ea10}
\wt P_{std}= \pmatrix{k - q r(Q,\wt P)\cr n k - n q r(Q,\wt P)\cr
2 k n q - n q^2 r(Q,\wt P) + a r(Q,\wt P) \cr r(Q,\wt P) \cr 0 \cr \cdot
\cr \cdot \cr 0}\, .
\ee
We choose $q$ such that $k -q \, r(Q,\wt P)$ is an integer between
0 and $r(Q,\wt P)-1$. By eq.\refb{ea3} this is a representative of
$-u_1(Q,\wt P)$ in the range $[0, r(Q,\wt P)-1]$. Hence it is
determined uniquely by $u_1(Q,\wt P)$. Let us call this
integer $d(Q,\wt P)$. We can then express \refb{ea10} as 
\be \label{ea1a}
\wt P_{std} = \pmatrix{d(Q,\wt P)\cr n d(Q,\wt P)\cr
b \cr r(Q,\wt P) \cr 0 \cr \cdot
\cr \cdot \cr 0}\, ,
\ee
where $b$ is a constant. It is determined by equating $(\wt
P_{std})^2$ to
$\wt P^2$:
\be \label{ea11}
2\, n \, d(Q,\wt P)^2 + 2\, b \, r(Q,\wt P) = \wt P^2\, .
\ee
Since $n = -Q^2/2$, this determines the form of $Q$ and
$\wt P_{std}$ completely in terms of the invariants
$Q^2$, $\wt P^2$, $r(Q,\wt P)$ and $d(Q,\wt P)$.  Thus
any two pairs of charge vectors $(Q_1,\wt P_1)$ and 
$(Q_2, \wt P_2)$ having same values of
these invariants and 
satisfying $Q_1\cdot \wt P_1=Q_2\cdot \wt P_2=0$ 
can be related to each other by a T-duaity transformation, 
since each pair can be brought by a
T-duality transformation to the standard form $(Q,\wt P_{std})$
given in \refb{ea1}, \refb{ea1a}. This is the desired result.

\sectiono{Analysis of $r(Q,P)=1$ condition} \label{sb}

In this appendix we shall 
derive a  condition on $Q^2$, $P^2$ and $Q\cdot P$
which is sufficient but not necessary to gurantee that
$r(Q,P)=1$.

As usual,
we shall assume that $Q$ and $P$ are primitive vectors of the
lattice. 
In this case we can represent $Q$ and $P$ as in \refb{einthis}.
This gives
\be \label{esb2}
Q^2 P^2 - (Q\cdot P)^2 = r(Q,P)^2 \left(e_1^2 e_2^2 
- (e_1\cdot e_2)^2\right)\, .
\ee
Thus in order for $r(Q,P)$ to be different from 1,  
$Q^2 P^2
- (Q\cdot P)^2$ must have a factor that is square of an integer.
Conversely, if $Q^2 P^2 - (Q\cdot P)^2$ is square free 
we can conclude that $r(Q,P)=1$.
In particular, for
$Q^2=P^2=-2$ and $Q\cdot P=\pm 1$
we have $Q^2 P^2 - (Q\cdot P)^2=3$. Since this is square
free we must have $r(Q,P)=1$.

So far we have taken $Q$ and $P$ to be arbitrary vectors in the
lattice. However if $Q$ and $P$ are to be identified as the
elements of the root lattice of a gauge algebra then the
induced metric on the vector space $E$ spanned by $Q$ and $P$
is euclidean. In this case
we can do slightly better by noting that since the lattice
is even, $e_1^2$ and $e_2^2$ must be even, while $e_1\cdot e_2$
is an integer. Thus $e_1^2 e_2^2$ is a multiple
of 4, while $(e_1\cdot e_2)^2$ has the form $4 s$ or $4s+1$
for some integer $s$. This implies that for positive $e_1^2 e_2^2
- (e_1\cdot e_2)^2$ -- which is the case since the induced metric in
$E$ is euclidean -- we must have
$e_1^2 e_2^2 - (e_1\cdot e_2)^2\ge 3$. Thus in order
for $r(Q,P)$ to be different from 1, the combination $Q^2 P^2 
-(Q\cdot P)^2$ must have the form $k l^2$ with $k\ge 3$, 
$l\ge 2$.

While the above analysis tells us under what condition $r(Q,P)=1$,
it does not tell us that if $Q^2 P^2 
-(Q\cdot P)^2$ has the form $k l^2$ with $k\ge 3$, $l\ge 2$ then
$r(Q,P)$ is necessarily larger than one. 
Thus $Q^2P^2 - (Q\cdot P)^2$ being square free is sufficient
but not necessary for $r(Q,P)$ to be 1.


\end{document}